\documentclass[
    ,draft            
  ]
  {aipproc}

\layoutstyle{6x9}

\usepackage{color}

\def\cp{$CP$}

\def\ra{\!\rightarrow\!}  
\def\bjpsiks{$B^0\ra J/\psi\,K^0_S$}
\def\bbar{\overline{B}{}^0}
\def\dbar{\overline{D}{}^0}

\def\gevm{~GeV/$c^2$}

\def\simge{\mathrel{%
   \rlap{\raise 0.511ex \hbox{$>$}}{\lower 0.511ex \hbox{$\sim$}}}}
\def\simle{\mathrel{
   \rlap{\raise 0.511ex \hbox{$<$}}{\lower 0.511ex \hbox{$\sim$}}}}

\begin{document}
\begin{flushright}
UCHEP-09-01
\end{flushright}
\vspace*{-0.34in}

\title{
Complementarity: Flavor Physics in the LHC Era}




\author{A. J. Schwartz}{
  address={Department of Physics, 
University of Cincinnati, P.O. Box 210011, Cincinnati, Ohio 45221}
}

\begin{abstract}
The LHC physics era is about to commence; here we discuss a
complementary physics program that would be realized at a high
luminosity flavor factory. A flavor factory experiment can 
search for new physics in \cp\ asymmetries, 
inclusive decay processes, rare leptonic processes, absolute
branching fractions, and other measurements that are challenging
or not feasible at the LHC. Such measurements would provide good 
sensitivity to new physics phases, the presence of a charged Higgs, 
and supersymmetric couplings. The charged Higgs mass range probed 
is similar to that accessible at the LHC.
\end{abstract}

\maketitle

\section{Introduction}

Particle physics is now entering the era of the LHC: the
experiments that will soon be running at this accelerator
are expected to shed light on the mechanism of electroweak
symmetry breaking and establish whether new particles and
interactions are present. However, the high center-of-mass 
energy utilized by the LHC is not the only method to access 
new physics (NP); this can also be done at lower energies
by studying decays of heavy flavors
($D$ and $B$ mesons, and $\tau$ leptons) with large
data sets. The latter type of experiment is known as a 
``flavor factory'' or often a ``$B$ Factory''
due to the prominent role played by $B$ meson decays
in understanding the Cabibbo-Kobayashi-Maskawa weak mixing
matrix. In the LHC era, there is a unique and complementary
role played by a flavor factory, and currently two groups 
are proposing to build such facilities: Belle-II~\cite{belleII} 
in Japan, which is an upgrade of the Belle experiment~\cite{belle},
and SuperB~\cite{superB} in Italy, which is an evolution of
the Babar experiment~\cite{babar}. These proposed experiments 
will nominally record 20-50 times the amount of data recorded 
by Belle and Babar, with the goal of addressing the following 
issues: are there more than three generations? Why are quark masses 
so different? Why is there such an unusual pattern of CKM weak
coupling strengths, and what causes the phase in the CKM matrix?
Is this the only phase present? Given the small amount of \cp\ 
violation observed in $K$ and $B$ decays, 
why is our universe overwhelmingly matter-dominated?

A flavor factory can search for NP in 
\cp\ asymmetries, inclusive decay processes, rare leptonic decays, 
absolute branching fractions, and other processes not easily 
accessible at the LHC. A flavor factory probes processes that 
occur at 1-loop in the Standard Model (SM) but may occur at tree 
level in NP scenarios; such SM-loop processes probe energy scales
that cannot be accessed directly at the LHC. If supersymmetry is
in fact observed at the LHC, a crucial question will be: how is
it broken? By making detailed studies of flavor couplings, a
flavor factory can address this.

\begin{figure}
\includegraphics[height=.25\textheight,angle=-90]{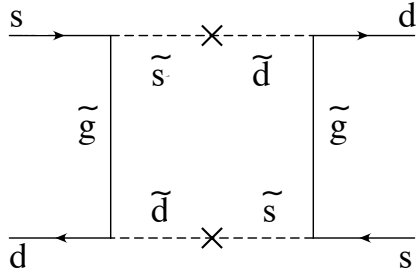}
\caption{A $\Delta F=2$ gluino-squark diagram.}
\label{fig:generic_loop}
\end{figure}

Most scenarios of new physics can be classified as either
supersymmetric theories,
little Higgs models,
models with extra dimensions,
left-right models, or
strongly coupled models~\cite{petrov}.
All of these have implications for flavor physics. For 
example, consider the $\Delta F\!=\!2$ gluino-squark diagram 
shown in Fig.~\ref{fig:generic_loop}. This diagram can give 
rise to neutral meson mixing. By dimensional considerations, 
the effective four-quark operators governing such mixing 
must have the form
$(C^{}_{NP}/\Lambda^2_{NP}) [\bar{d}^{}_{j L}\gamma^{}_\mu d^{}_{iL}]^2$
plus other possible chirality structure, where $\Lambda^{}_{NP}$ 
is the energy scale associated with NP, and $C^{}_{NP}$ is a 
NP Wilson coefficient of $O(1)$. The corresponding term in 
the SM, e.g., for $B^0$-$\bbar$ mixing, is
\begin{eqnarray}
\left(\frac{g^2}{8\pi M^{}_W}\right)^2 (V^*_{td}V^{}_{tb})^2\,
[\bar{d}^{}_{L}\gamma^{}_\mu b^{}_{L}]^2 & \approx & 
\frac{(\sin^3\theta^{}_c)^2}{(2\times 2.5~{\rm TeV})^2}\,
[\bar{d}^{}_{L}\gamma^{}_\mu b^{}_{L}]^2\,.
\label{eqn:np}
\end{eqnarray}
Since we do not yet observe NP effects, the NP operator must 
be less  than the SM term~\eqref{eqn:np}, or equivalently
$\Lambda^{}_{NP}>
\sqrt{C^{}_{NP}\cdot 4\cdot(2.5~{\rm TeV})^2/(\sin^3\theta^{}_c)^2}
\sim 400$~TeV. The fact that this scale is much larger than the 
weak scale is a manifestation of the ``flavor problem:''
new physics needed at the TeV scale to stabilize the electroweak
symmetry-breaking scale must have a highly non-generic flavor 
structure.

For most NP models, the coupling strengths across generations 
are not predicted. If one assumes that the only source of 
flavor-changing neutral currents (FCNC) is the Yukawa couplings 
--\,and there is no {\it a priori\/} reason why this 
should be so\,-- then the FCNC are CKM-suppressed. 
For example, for $B^0$-$\bbar$ mixing
\begin{eqnarray}
H^{\rm eff}_{NP} & \approx &
\frac{C^{}_{NP}}{\Lambda^2_{NP}} 
(V^*_{td}V^{}_{tb})^2\,[\bar{d}^{}_{L}\gamma^{}_\mu b^{}_{L}]^2\,,
\label{eqn:mfv}
\end{eqnarray}
which implies
$\Lambda^{}_{NP}>
\sqrt{C^{}_{NP}\cdot 4\cdot(2.5~{\rm TeV})^2}\sim 5$~TeV. 
This scenario is known as ``Minimal Flavor Violation.'' 
Although it substantially lowers the scale at which NP 
could appear, the scale remains higher than 
what the LHC can comfortably probe.

In the remainder of this paper we discuss five types 
of measurements for which a flavor factory has excellent 
potential for identifying NP if present:
{\it (a)\/} measuring mixing phases;
{\it (b)\/} measuring decay phases;
{\it (c)\/} detecting a charged Higgs via leptonic decays;
{\it (d)\/} identifying new physics via $b\ra s$ transitions; and 
{\it (e)\/} determining how supersymmetry is broken via measurements 
of $\sin 2\phi^{}_1$ (a combination of weak and decay phases).
Here we limit our discussions to $B$ decays, but a flavor 
factory also produces prodigious amounts of $D$ and $\tau^\pm$
decays and will also use these to search for~NP.

\section{New Physics in Mixing Phases}

A flavor factory can measure weak mixing phases, which is
a phase that enters a $B^0$-$\bbar$ ``oscillation'' amplitude. 
In the SM such amplitudes consist of $\Delta B=2$ loop diagrams 
and thus are especially sensitive to NP.
The original goal of Belle and Babar was to measure the phase
$\phi^{}_1\equiv {\rm Arg}(-V^{*}_{cb}V^{}_{cd}/V^{*}_{tb}V^{}_{td})$,
whose size is mainly determined by the mixing phase
${\rm Arg}(V^{*}_{tb}V^{}_{td}$). The method is as
follows: a $B^0$ oscillates to a $\bbar$ and subsequently
decays to a self-conjugate final state that a $B^0$ can
decay to directly. This amplitude (due to mixing), as 
compared to the direct amplitude, contains an additional 
weak phase (that of mixing). The two amplitudes 
interfere, and the interference term in the decay rate 
is proportional to the sine of twice the overall weak phase. 
Thus, measuring the decay time dependence allows one to 
fit for the interference term and determine this weak phase.

\begin{figure}
\includegraphics[height=.45\textheight]{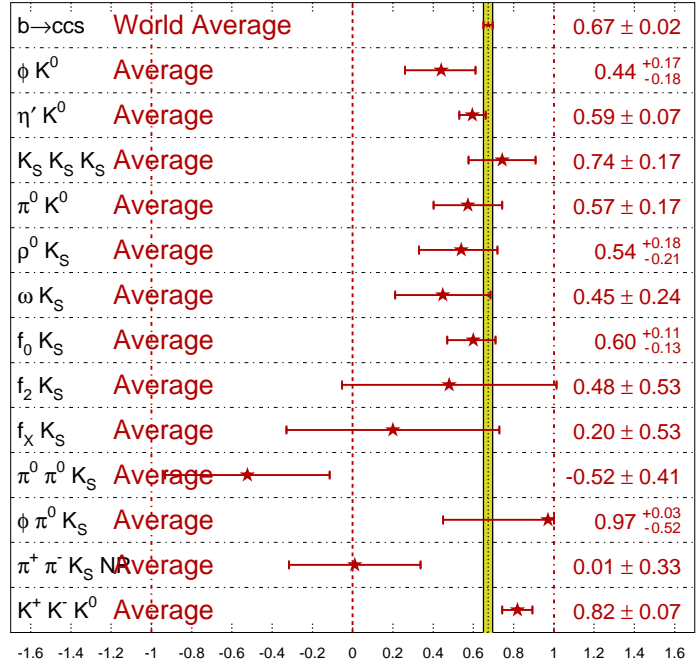}
\caption{World average values of $\sin 2\phi^{}_1$ measured
using various decay modes (left), from the Heavy Flavor Averaging
Group (HFAG)~\cite{hfag:sin2phi1}.
The $b\ra c\bar{c}s$ value is dominated by \bjpsiks\ decays.}
\label{fig:weakphase1}
\end{figure}

\begin{figure}
\vbox{
\includegraphics[height=.36\textheight]{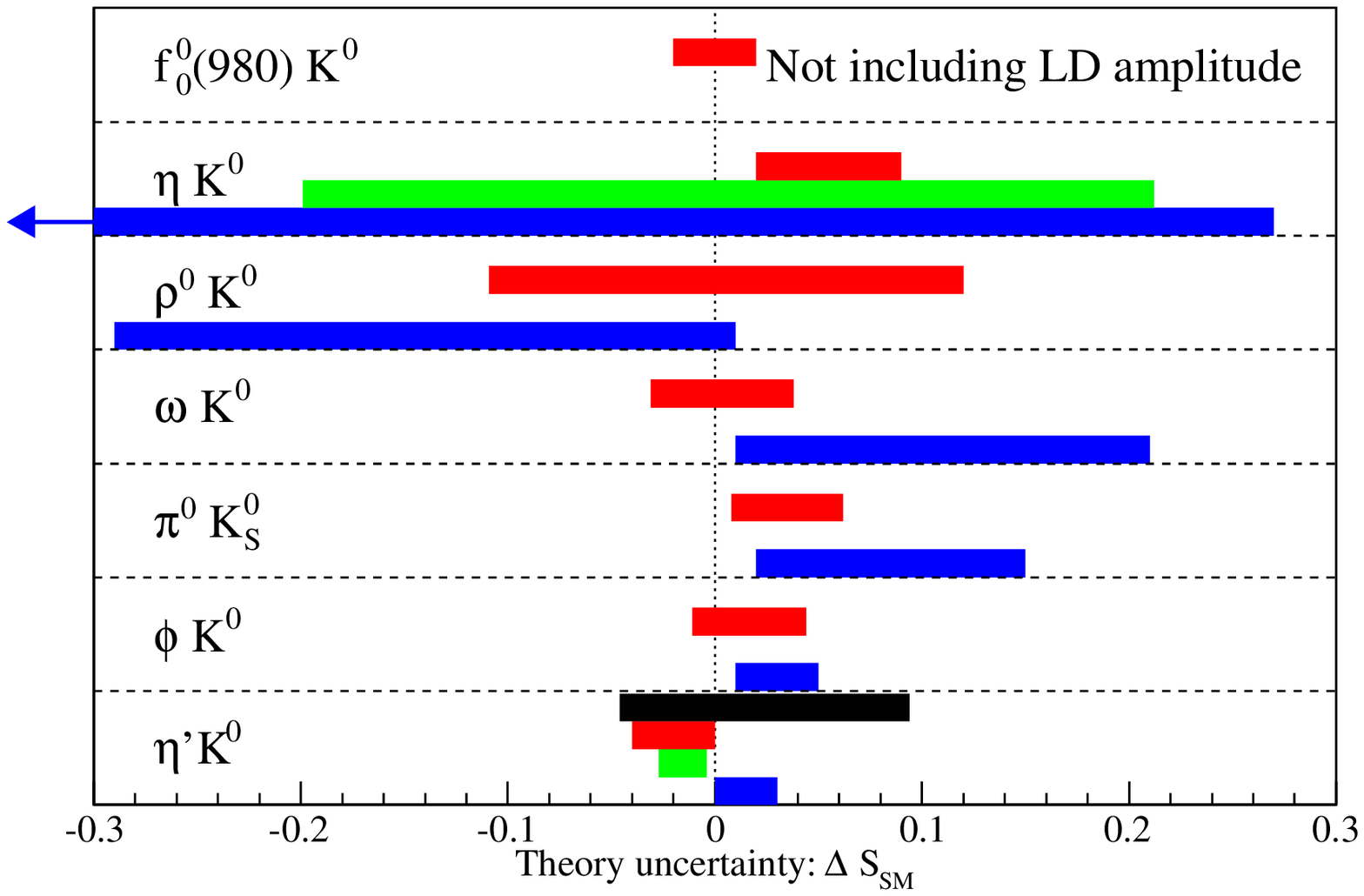}
\vskip-0.05in\hspace*{0.70in}
\includegraphics[height=.40\textheight,angle=-90]{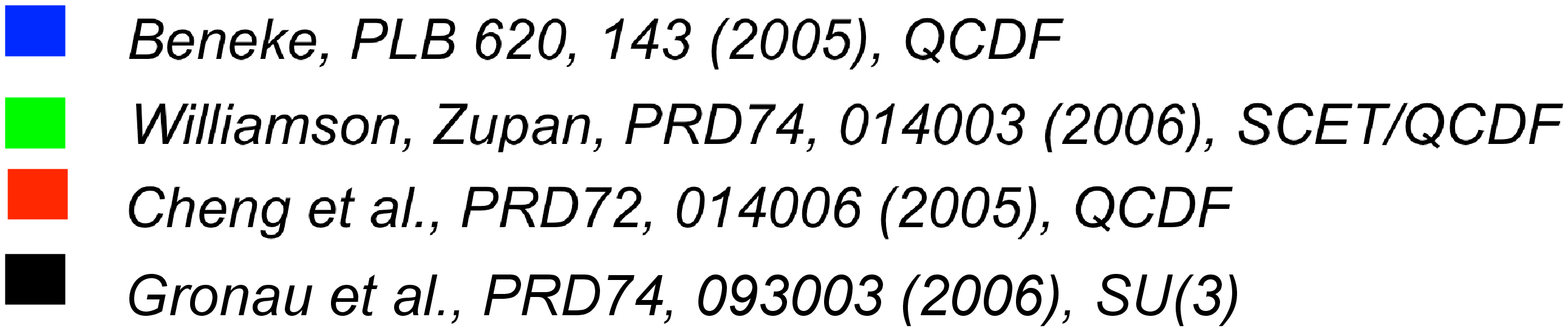}
}
\caption{Compilation of theoretical predictions for 
$\Delta S\equiv(\sin 2\phi^{}_1)_{q\bar{q}s} - (\sin 2\phi^{}_1)_{c\bar{c}s}$,
from Ref.~\cite{bevan}.}
\label{fig:weakphase2}
\end{figure}

Applying the method to high-statistics \bjpsiks\ and
other $b\ra c\bar{c}s$ decay channels measures the phase 
$\phi^{}_1$; the result is 
$\sin 2\phi^{}_1 = 0.670\,\pm 0.023$~\cite{hfag:sin2phi1}. 
This method has been extended to $b\ra ss\bar{s}$ and 
$b\ra sd\bar{d}$ penguin decay modes, which should have 
the same overall weak phase~($\phi^{}_1$) up to small 
corrections of 
$O(\sin^2\theta^{}_C)$. The results are tabulated in 
Fig.~\ref{fig:weakphase1}; also shown for comparison 
is the world average value measured from $b\ra c\bar{c}s$ 
decays. The table shows a systematic shift for 
$b\ra sq\bar{q}$ to lower values, although the 
current statistical errors preclude drawing a 
firm conclusion. However, most theoretical 
predictions prefer a shift to {\it higher\/} values 
(see Fig.~\ref{fig:weakphase2});
this difference could be a sign of a new non-SM phase.
A future flavor factory should clarify this, as the
expected statistical improvement in measuring 
$\sin 2\phi^{}_1$ from $b\ra sq\bar{q}$ decays
is a factor of~5--10.

\section{New Physics in Decay Phases}

A flavor factory can also search for new physics (NP) 
in {\it decay\/} phases, i.e., phases that appear directly
in a $b\ra q$ decay amplitude (no $B^0$-$\bbar$ mixing is
needed). Decay phases give rise to direct \cp\ violation,
i.e., a difference between 
$\Gamma(B\ra f)$ and $\Gamma(\overline{B}\ra\bar{f})$
due to interference between two or more decay amplitudes
(e.g., penguin and tree) with different weak phases.
Explicitly, if $B\ra f$ proceeds via a tree amplitude
${\cal A}^{}_t = |A^{}_t| e^{i(\phi^{}_t + \delta^{}_t)}$ 
and a penguin amplitude 
${\cal A}^{}_p = |A^{}_p| e^{i(\phi^{}_p + \delta^{}_p)}$, 
where $\phi$ and $\delta$ are weak and strong phases,
respectively,
then $\overline{B}\ra\bar{f}$ proceeds via
$|A^{}_t| e^{i(-\phi^{}_t + \delta^{}_t)}$ and 
$|A^{}_p| e^{i(-\phi^{}_p + \delta^{}_p)}$. Summing 
the amplitudes and squaring gives decay rates
$R=|A^{}_t|^2 + |A^{}_p|^2 + 
4|A^{}_t||A^{}_p|\cos(\Delta\delta\pm\Delta\phi)$, 
where $\Delta\delta = \delta^{}_t-\delta^{}_p$ and
$\Delta\phi = \phi^{}_t-\phi^{}_p$. Taking the sums 
and differences of the decay rates gives
\begin{eqnarray}
A^{}_{CP} & \equiv & \frac{\Gamma(B\ra f)-\Gamma(\overline{B}\ra\bar{f})}
{\Gamma(B\ra f)+\Gamma(\overline{B}\ra\bar{f})}\ \propto\ 
\sin \Delta\phi\,\sin \Delta\delta\,.
\end{eqnarray}
Thus a direct \cp\ asymmetry arises only if the two 
contributing decay amplitudes have different weak 
($\phi$) and strong ($\delta$) phases.

\begin{figure}
\includegraphics[height=.50\textheight,angle=-90]{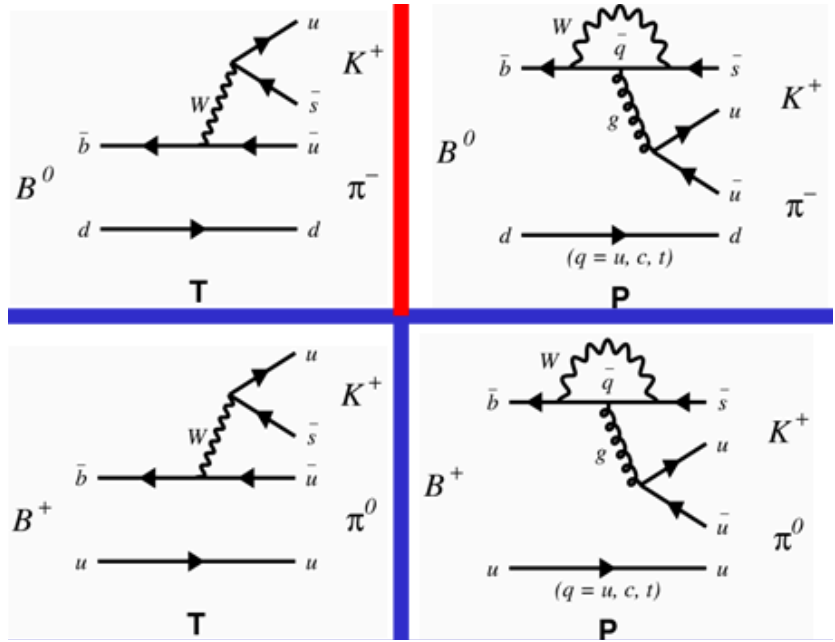}
\caption{
Tree amplitude (upper left) and penguin amplitude (upper right) 
contribution to $B^0\ra K^+\pi^-$;
tree amplitude (lower left) and penguin amplitude (lower right) 
contribution to $B^+\ra K^+\pi^0$.}
\label{fig:kpi}
\end{figure}

The charmless decays $B\ra K\pi$ proceed via both tree 
and penguin amplitudes and, because the final states are 
self-tagging, are well-suited for measuring $A^{}_{CP}$. The 
two amplitudes for the neutral decay $B^0\ra K^+\pi^-$ are shown
in Figs.~\ref{fig:kpi}a and \ref{fig:kpi}b; the measured 
\cp\ asymmetry is $-0.098\,^{+0.012}_{-0.011}$~\cite{hfag:acp}. 
The analogous diagrams for the charged decay $B^+\ra K^+\pi^0$ 
are shown in Figs.~\ref{fig:kpi}c and \ref{fig:kpi}d; as these
have the same weak and strong phases as those for $B^0$ decay, 
$A^{}_{CP}$ should be the same. However, the result for 
$B^+$ is $5.3\sigma$ away: $A^{}_{CP}=0.050\pm 0.025$~\cite{hfag:acp}. 
This discrepancy may indicate the presence of a new phase 
entering one of the decay amplitudes, e.g., in an NP-enhanced 
electroweak penguin~\cite{electr_peng}.
A future flavor factory can clarify this.

\section{Detecting a Charged Higgs}

A flavor factory in fact has similar sensitivity to a charged 
Higgs boson as does the LHC, but in complementary search channels.
A charged Higgs would manifest itself in inclusive $b\ra s\gamma$
decays, and in exclusive $B^+\ra\tau^+\nu$ and $B^0\ra D^{*-}\tau^+\nu$ 
decays. Although these final states are challenging to reconstruct 
due to missing particles, both Belle and Babar have reconstructed 
first samples~\cite{btaunu_belle,btotau_all} and a flavor factory 
would increase these substantially.

\begin{figure}
\hbox{
\includegraphics[height=.17\textheight,angle=-90]{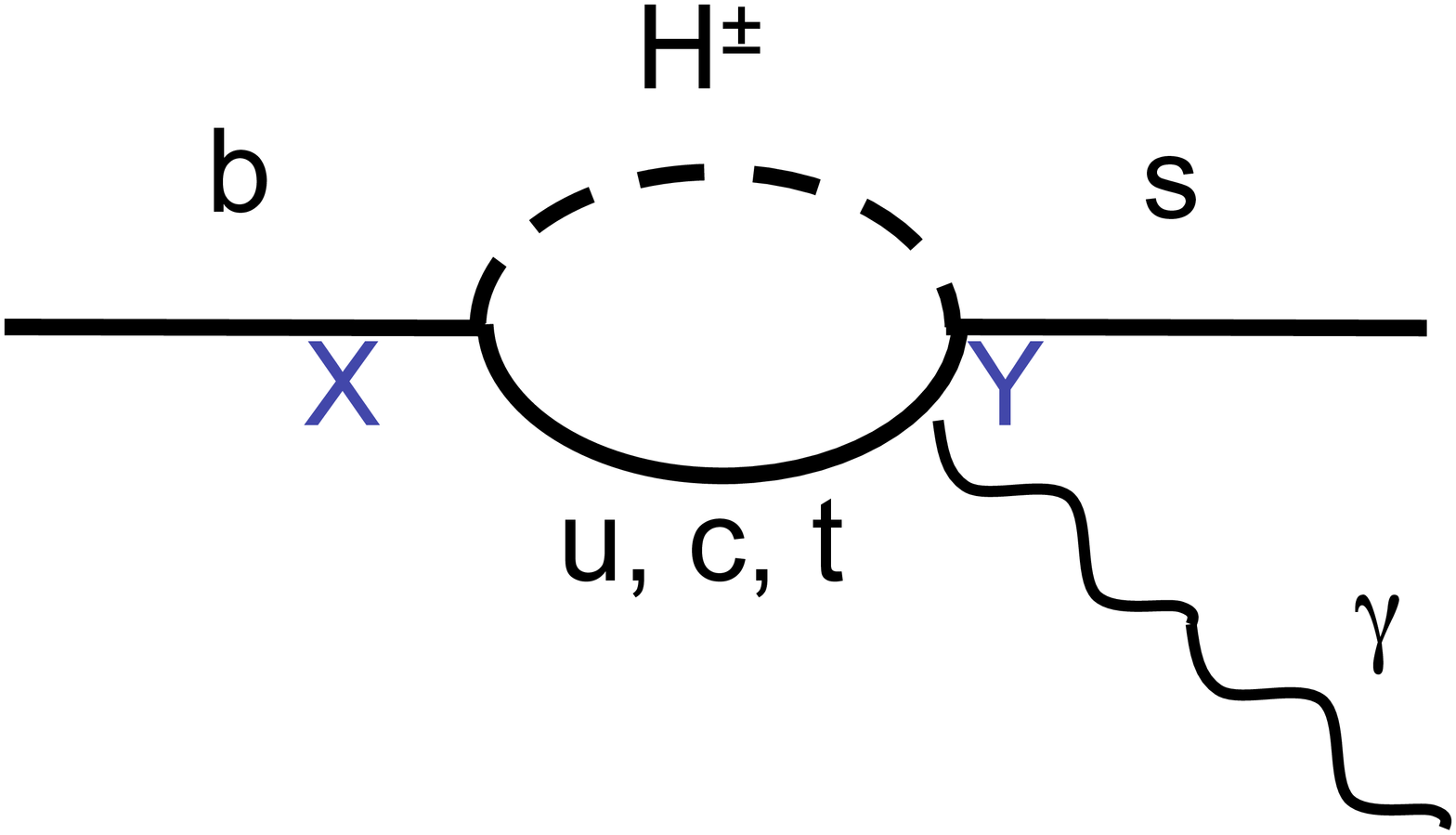}
\hskip0.40in
\includegraphics[height=.17\textheight,angle=-90]{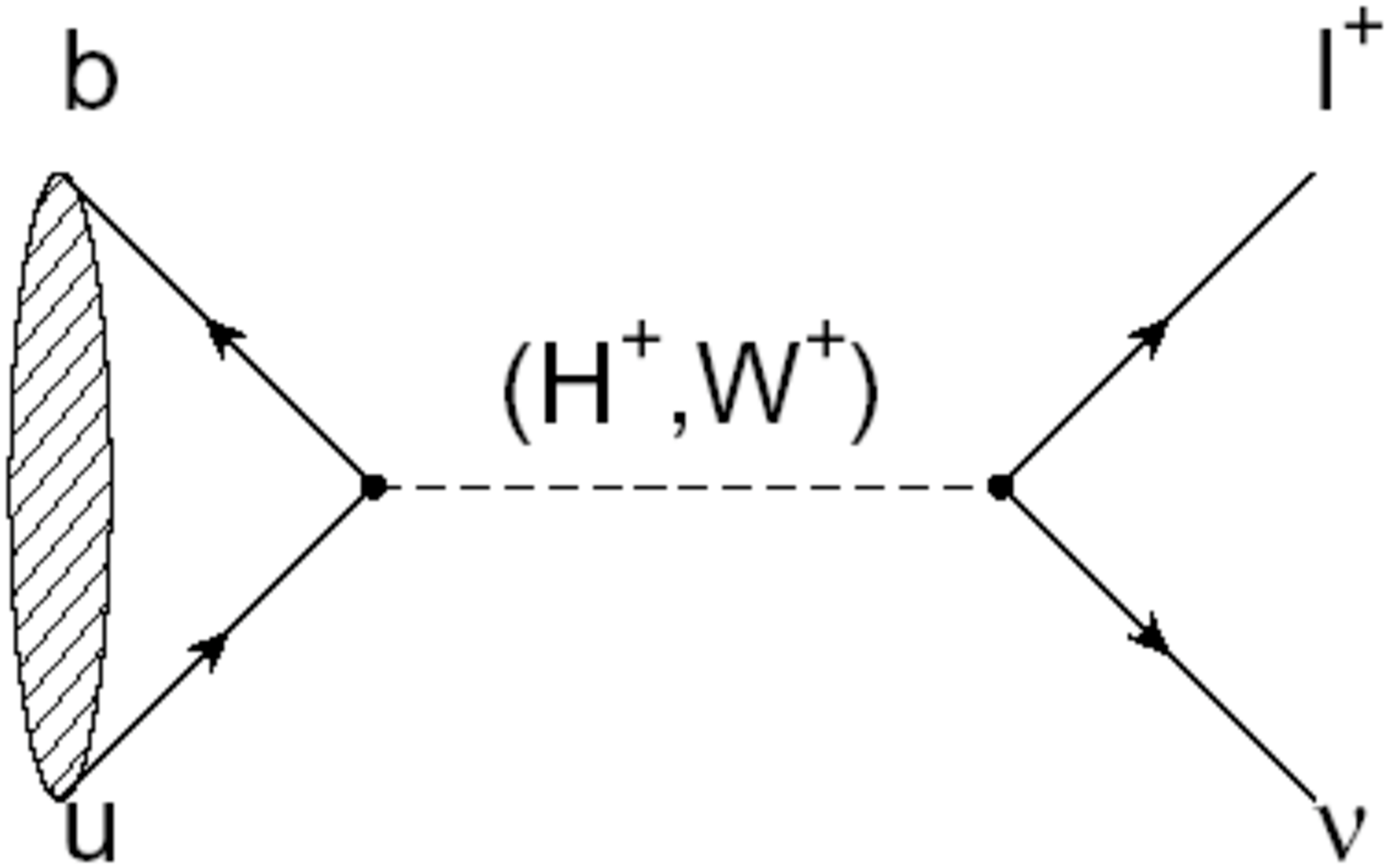}
\hskip0.40in
\includegraphics[height=.19\textheight,angle=-90]{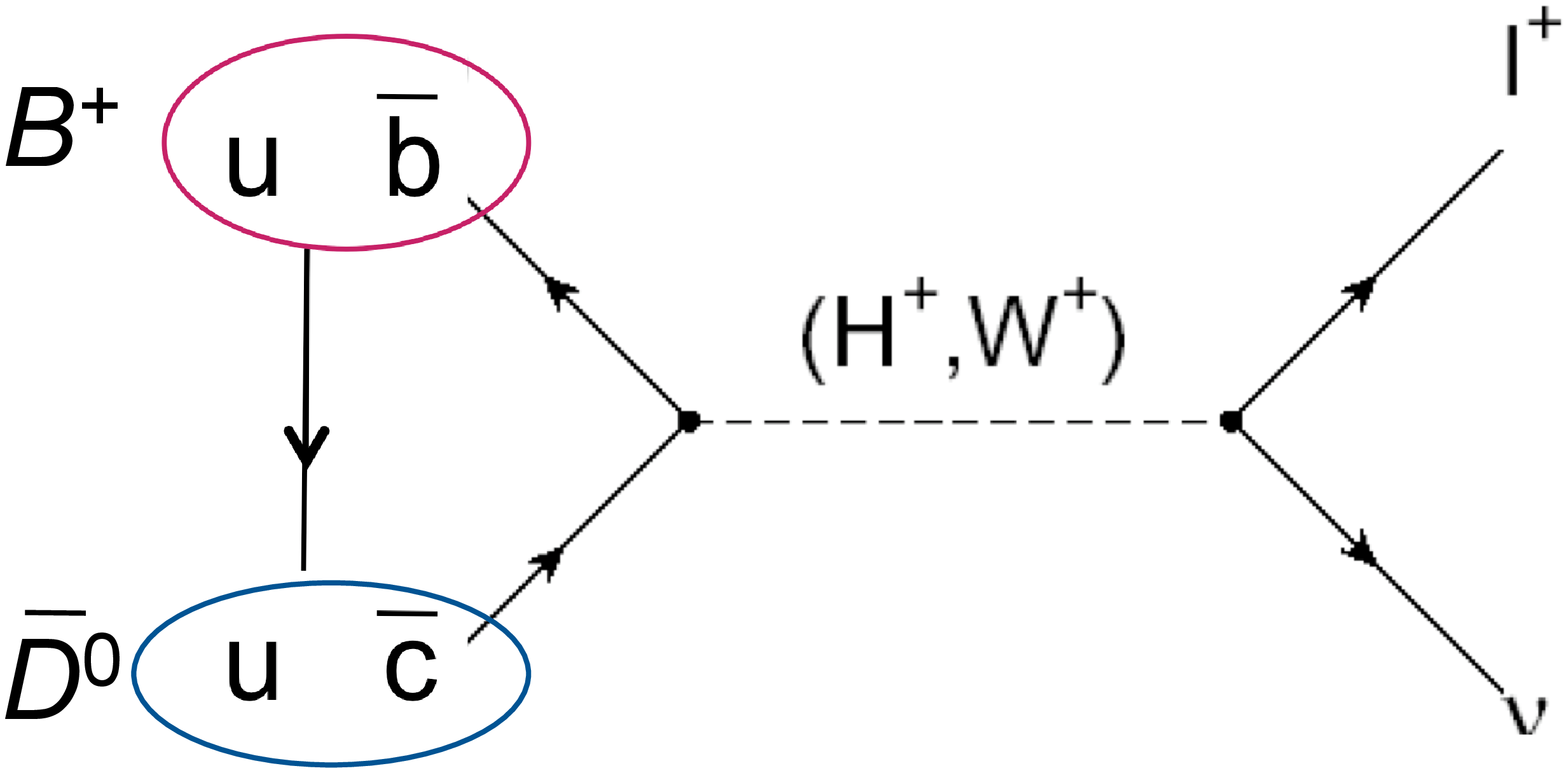}
}
\caption{Charged-Higgs-mediated diagrams for $b\ra s\,\gamma$ (left), 
$B^+\ra\tau^+\nu$ (middle), and $B^+\ra\overline{D}{}^{(*)0}\tau^+\nu$ (right).}
\label{fig:diagrams}
\end{figure}

\begin{figure}
\includegraphics[height=.38\textheight]{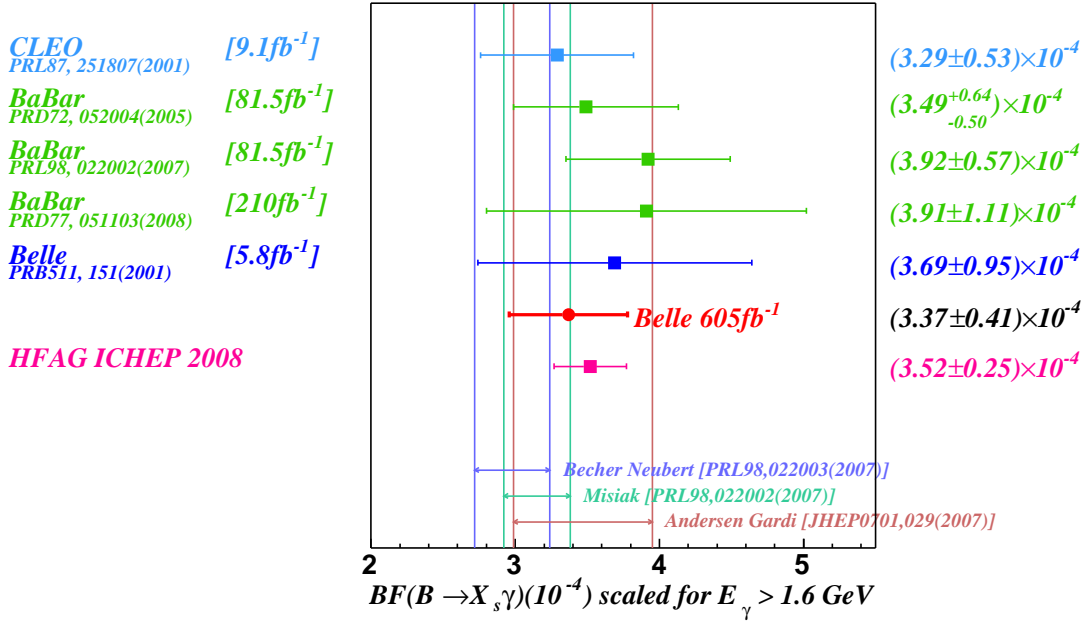}
\vspace*{0.50in}
\caption{Compilation of measurements of the inclusive process
$b\ra s\gamma$.
}
\label{fig:bsgamma_data}
\end{figure}

\begin{figure}
\includegraphics[height=.42\textheight,angle=-90]{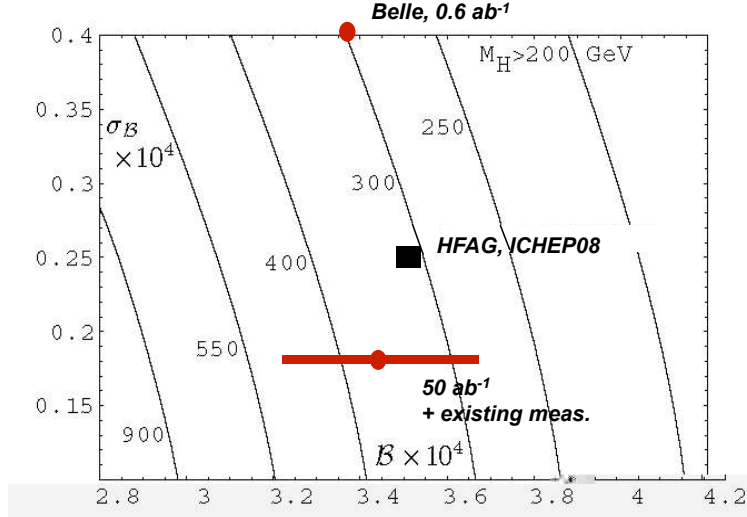}
\caption{95\% C.L. lower limits on $M^{}_{H^+}$ as a function of
the central value and error for ${\cal B}(b\ra s\gamma)$, from
Ref.~\cite{bsgamma_plot}.}
\label{fig:bsgamma_theory}
\end{figure}

For inclusive $b\ra s\gamma$ decays, one requires the presence 
of a high momentum photon and a charged kaon. The 
charged-Higgs-mediated diagram for this decay is shown 
in Fig.~\ref{fig:diagrams}a. Various measurements are 
tabulated in Fig.~\ref{fig:bsgamma_data}; the world 
average (WA) branching fraction is 
$(3.52\pm 0.25)\times 10^{-4}$~\cite{hfag:bsgamma}. 
An upper limit on the branching 
fraction sets a lower limit on $M^{}_H$; this is illustrated
in Fig.~\ref{fig:bsgamma_theory}, which plots the limit as 
a function of the branching fraction central value and error. 
The plot corresponds to all values of $\tan\beta$, where
$\beta$ is the ratio of vacuum expectation values for 
the two Higgs fields and is an unknown model parameter. 
From the figure one reads that the current WA branching 
fraction corresponds to a 95\% C.L. lower bound 
$M^{}_{H^+}>300$\gevm. At a future flavor factory,
this bound may increase to more then 500\gevm,
depending on the branching fraction central value.

\begin{figure}
\hbox{
\includegraphics[height=.35\textheight]{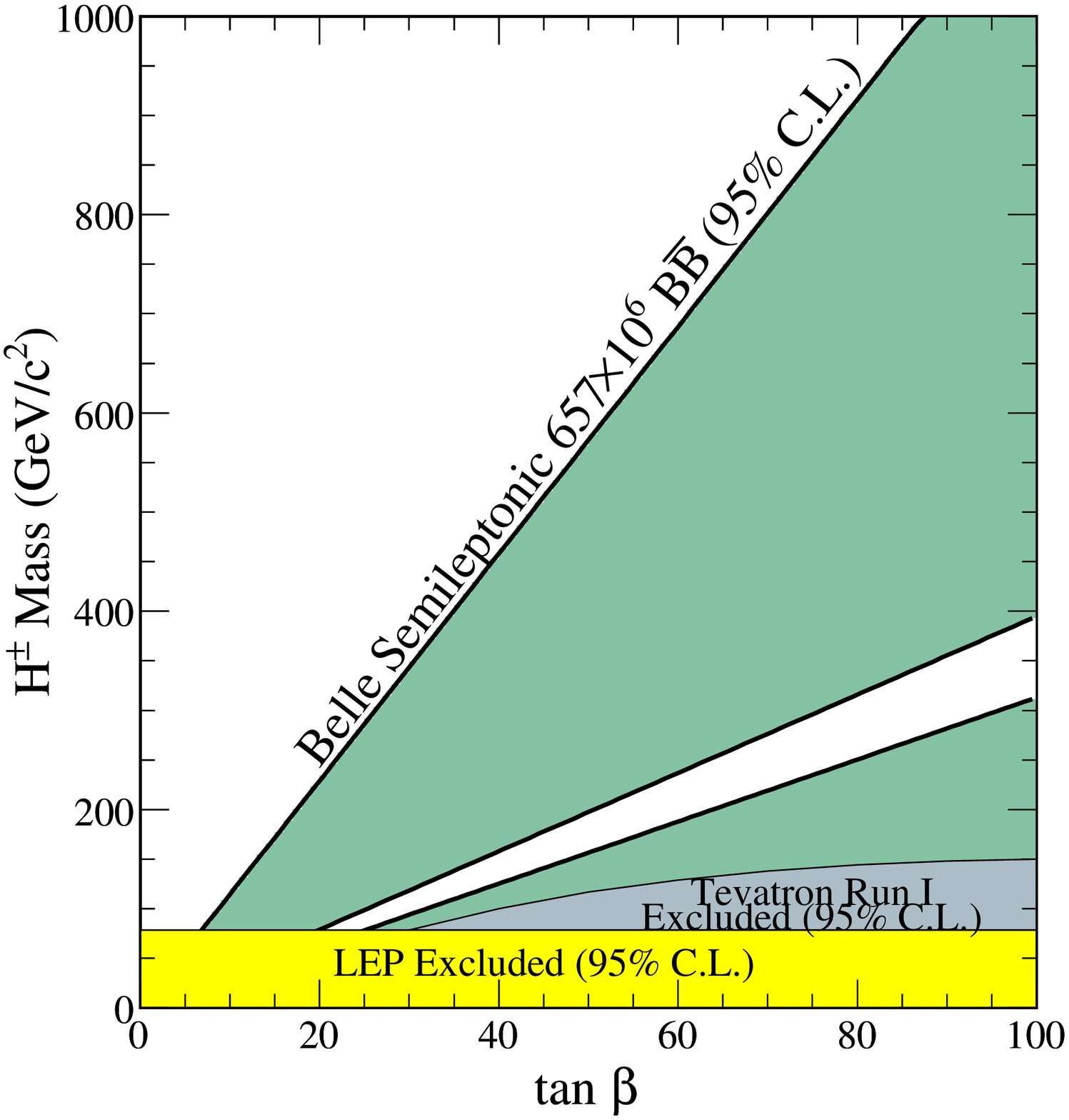}
\hskip0.20in
\includegraphics[height=.35\textheight]{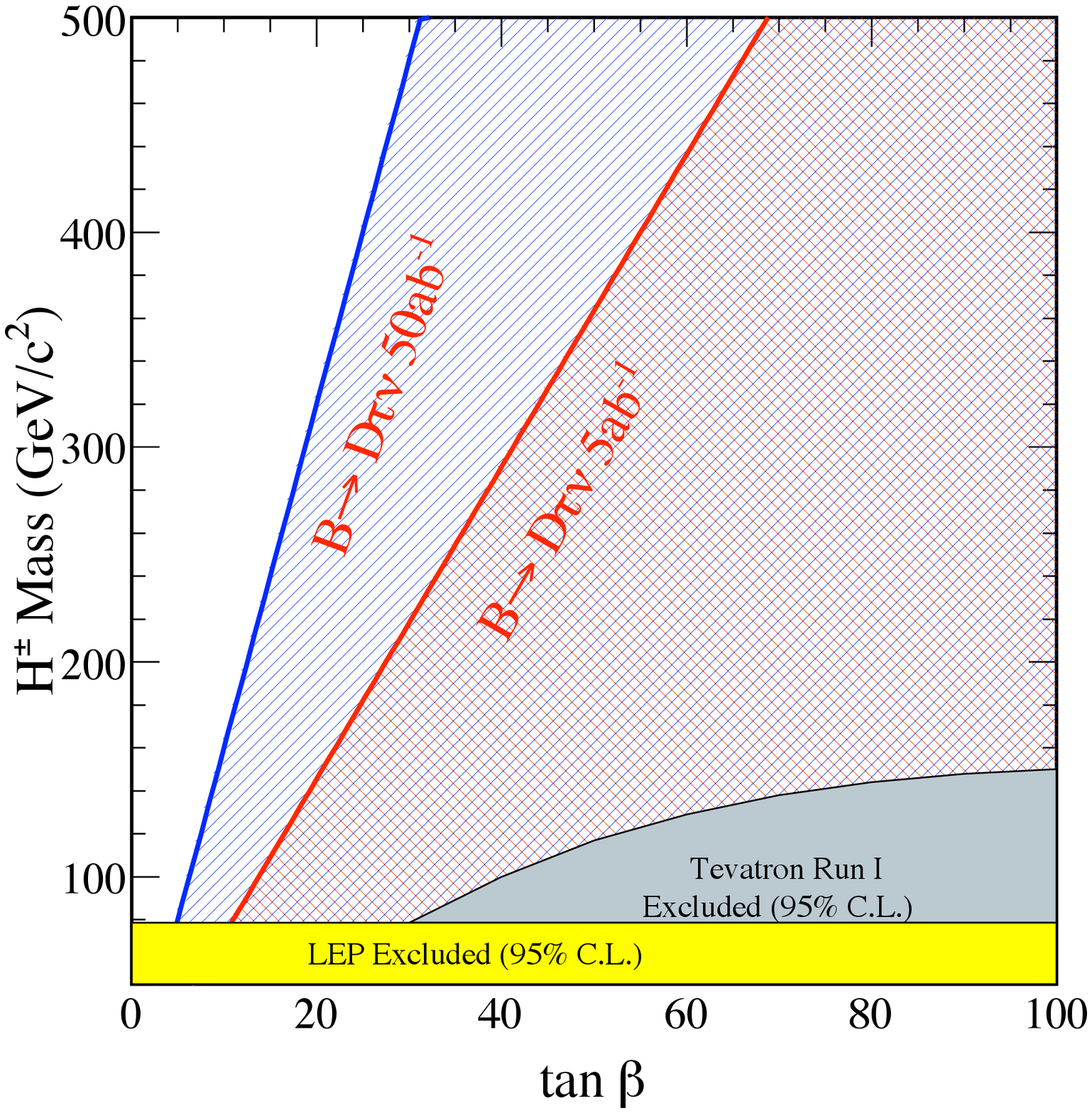}
}
\caption{
Excluded values of $M^{}_{H^+}$ (shaded regions) as a 
function of $\tan\beta$, from the upper limits on 
${\cal B}(B^+\ra\tau^+\nu)$ (left)~\cite{btaunu_belle_plot} 
and ${\cal B}(B^0\ra D^{*-}\tau^+\nu)$ (right)~\cite{cern_wg_epjc}.
In the right-most plot, the red (blue) shaded region corresponds
to 5~ab$^{-1}$ (50~ab$^{-1}$) of data.}
\label{fig:higgs_vs_tanb}
\end{figure}

A charged Higgs also affects $B^+\ra\tau^+\nu$ decays; 
the relevant diagram is shown in Fig.~\ref{fig:diagrams}b.
The analysis is challenging because the signal side has 
two missing neutrinos and thus cannot be fully reconstructed. 
The Belle result is 
${\cal B}(B^+\ra\tau^+\nu)=
(1.65\,^{+0.38}_{-0.37}\,^{+0.35}_{-0.37})\times 10^{-4}$~\cite{btaunu_belle}; 
dividing by the SM prediction 
${\cal B}^{}_{SM}=(0.796\,^{+0.154}_{-0.093})\times 10^{-4}$~\cite{btaunu_sm} 
yields
$r^{}_H\equiv {\cal B}^{}_{\rm measured}/{\cal B}^{}_{\rm SM}=
2.07\pm 0.74$~\cite{btaunu_belle_plot}.
Theoretically, in a two-Higgs doublet model 
$r^{}_H = (1-M^2_B\,\tan^2\beta/M^2_H)^2$~\cite{btaunu_theory};
an upper limit on $r^{}_H$ thus gives a lower limit on $M^{}_H$ 
that depends on $\tan\beta$. The result is shown in 
Fig.~\ref{fig:higgs_vs_tanb}a, where the shaded region 
is excluded. The allowed region (light band) in the middle 
of the shaded region results from a dip in $r^{}_H$ near 
$(\tan\beta/M^{}_H)\approx 1/M^{}_B$. This ``gap'' will be closed 
by measurements at a flavor factory of $B\ra D^{*}\tau^+\nu$ 
decays. The amplitude for this is governed by the $b\ra c$ 
transition shown in Fig.~\ref{fig:diagrams}c. The expected 
sensitivity of a flavor factory for 5~ab$^{-1}$ and 50~ab$^{-1}$ 
of data is shown in Fig.~\ref{fig:higgs_vs_tanb}b; for large 
$\tan\beta$, most of the $M^{}_H$ range accessible to the LHC 
is covered~\cite{lhc_sens}.

\vskip0.30in

\section{Identifying Supersymmetry}

Finally, we show that a flavor factory has suprisingly good
sensitivity to supersymmetry. If the LHC observes signs of 
supersymmetry, measurements from a flavor factory could prove 
crucial for distinguishing among various theoretical models 
and determining the mechanism by which supersymmetry is broken. 

\begin{figure}
\vbox{
\hbox{
\includegraphics[height=.28\textheight]{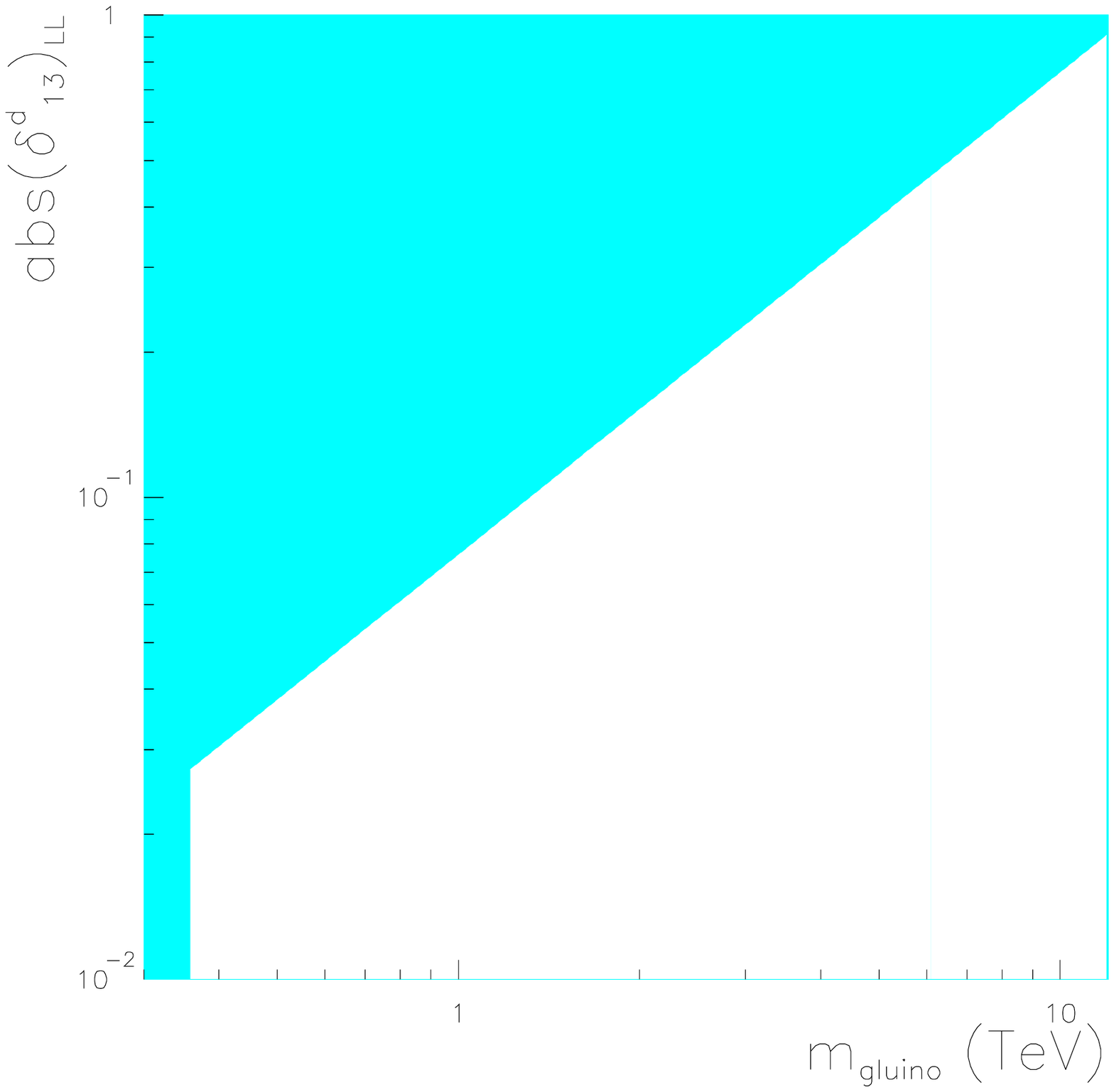}
\hskip0.30in
\includegraphics[height=.28\textheight]{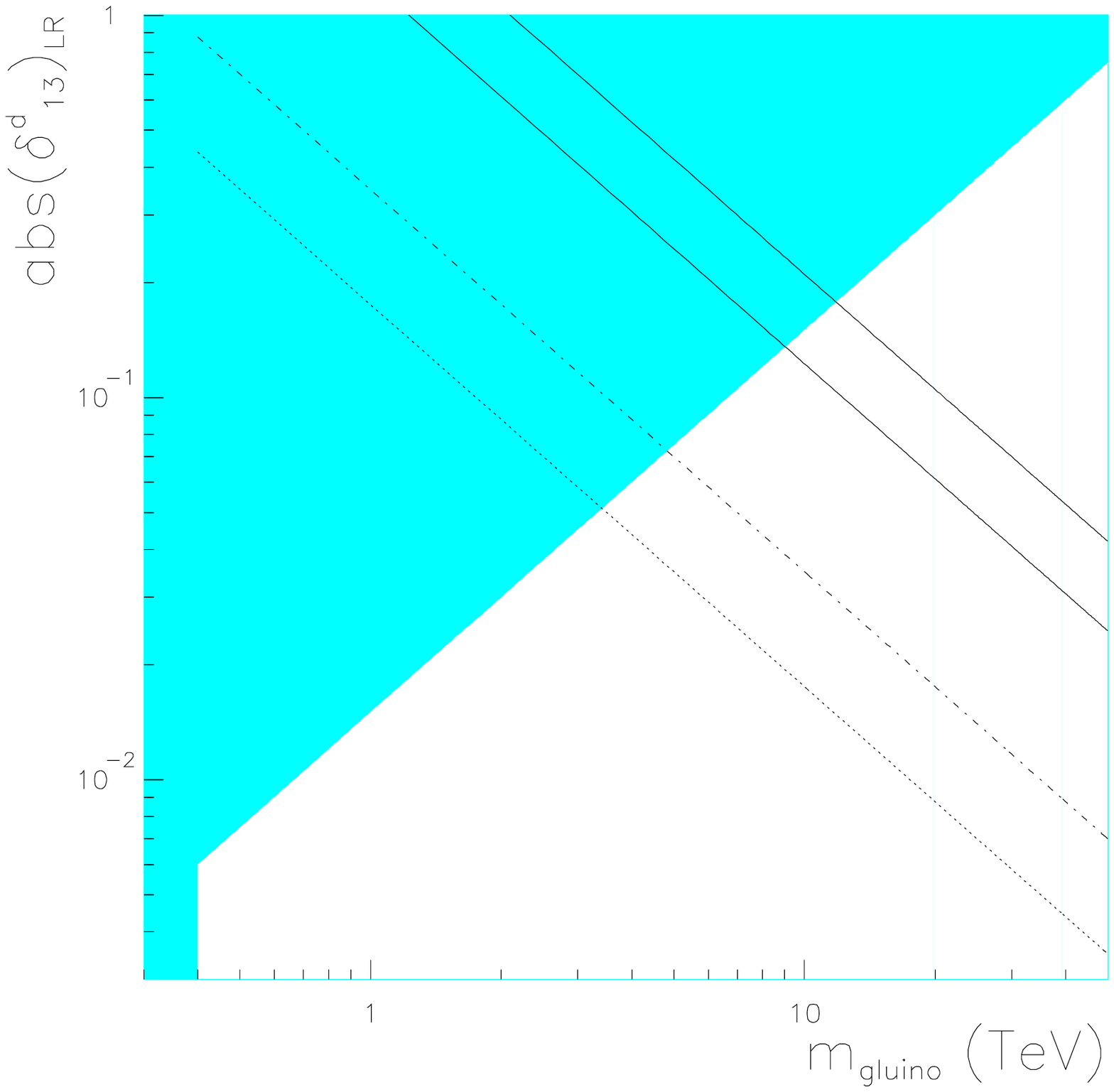}}
\vskip0.20in
\hbox{
\includegraphics[height=.28\textheight]{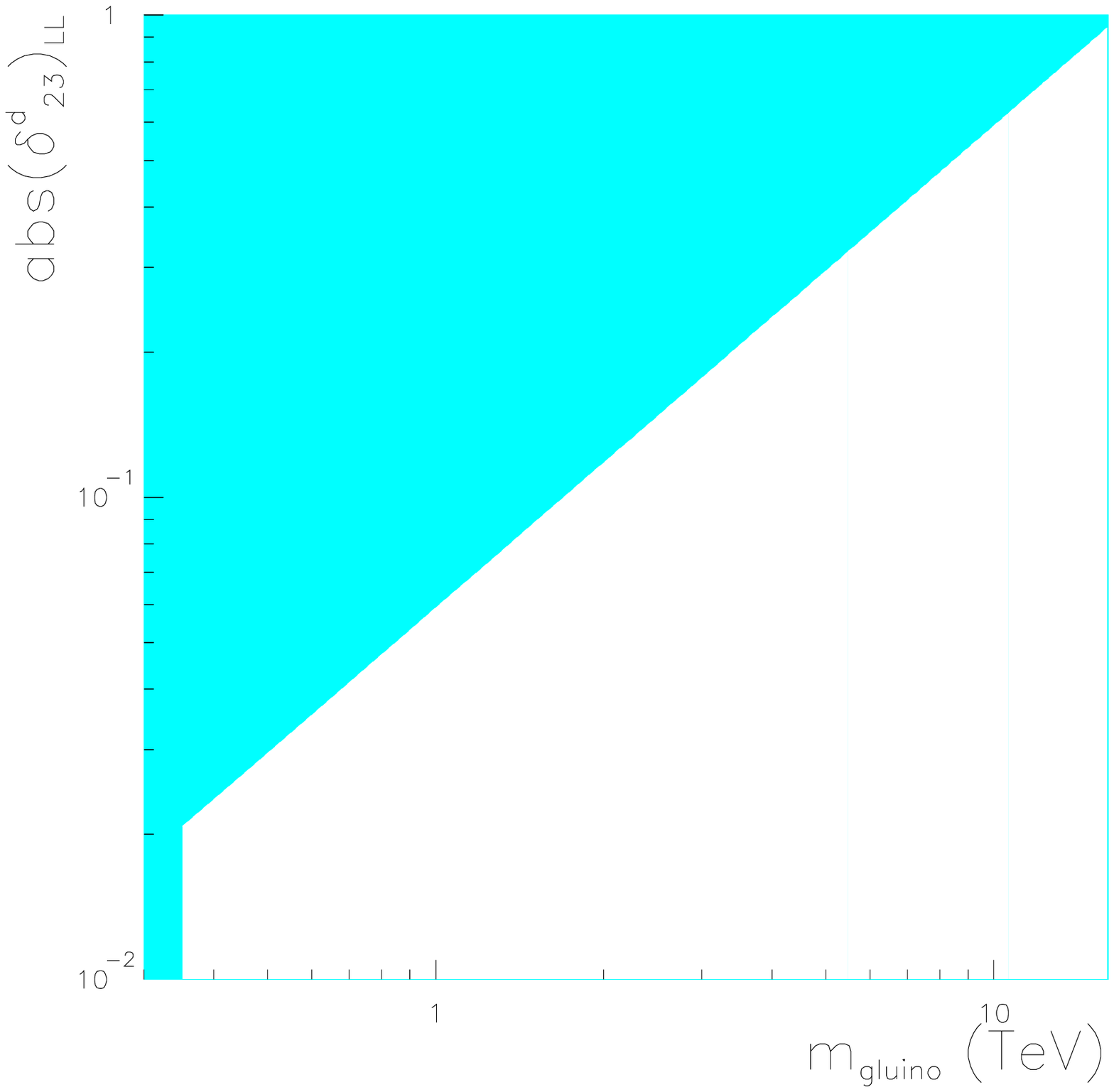}
\hskip0.30in
\includegraphics[height=.28\textheight]{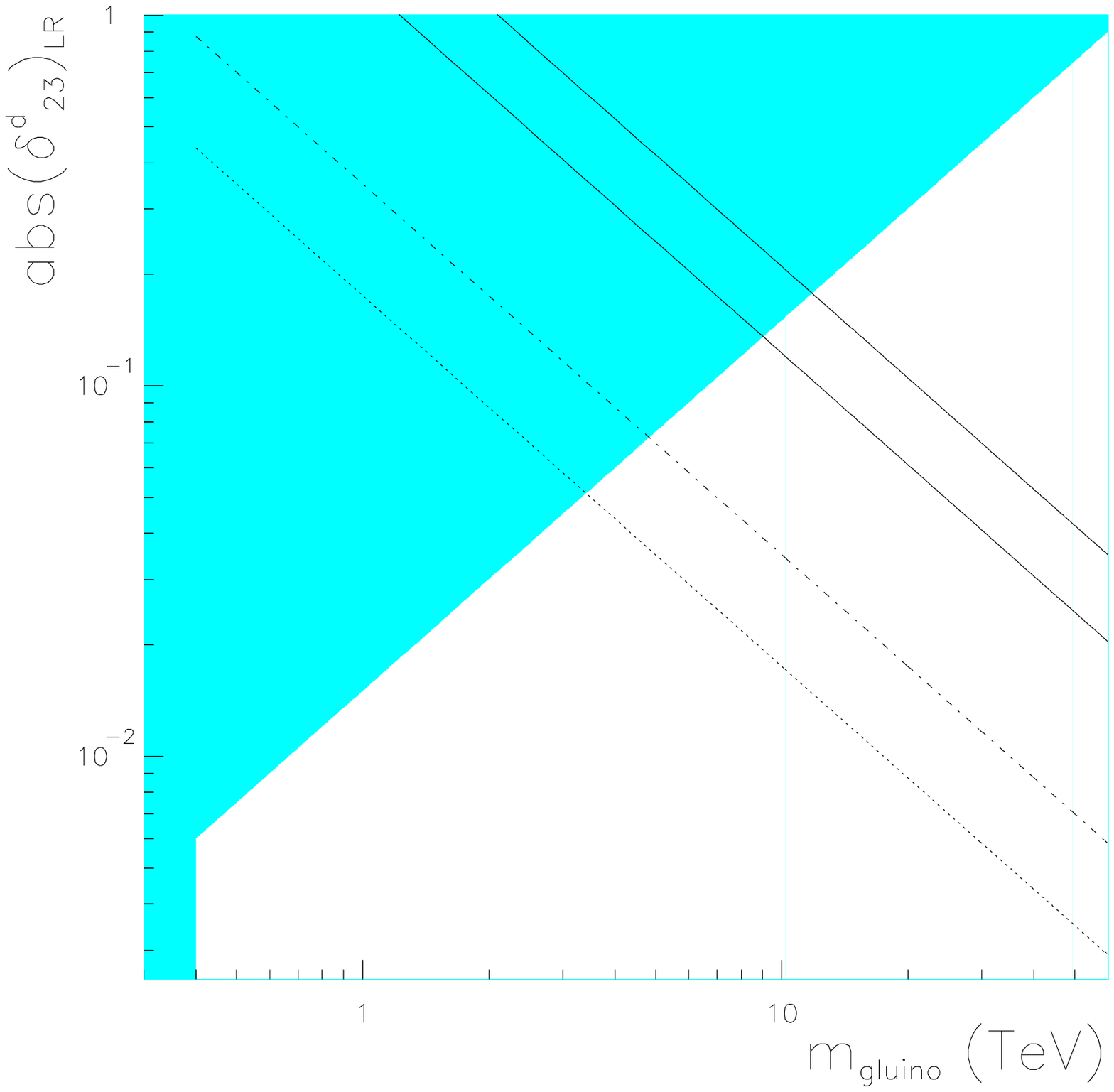}}
}
\caption{Values of mass insertion terms that would be measured 
non-zero with at least $3\sigma$ statistical significance 
(shaded regions), for 50~ab$^{-1}$ of data~\cite{cern_wg_arxiv}. 
The upper left plot is for $(\delta^{}_{13})^{}_{\rm LL}$;
the upper right plot for $(\delta^{}_{13})^{}_{\rm LR}$;
the lower left plot for $(\delta^{}_{23})^{}_{\rm LL}$; and
the lower right plot for $(\delta^{}_{23})^{}_{\rm LR}$. }
\label{fig:supersymmetry1}
\end{figure}

Supersymmetric theories are challenging to experimentally
confirm or exclude as there are many parameters to tune. 
For example, one way of calculating 
flavor-changing neutral-current (FCNC) processes in MSSM is 
to parameterize squark mass matrices with flavor-off-diagonal 
mass insertion terms. 
The corresponding sparticles can mediate SM-suppressed 
FCNC transitions such as $b\ra s$. Thus, measuring 
$b\ra s$ observables such as
${\cal B}(b\ra s\gamma)$, 
$A^{}_{CP}(b\ra s\gamma)$, 
${\cal B}(b\ra s\ell^+\ell^-)$, and
$A^{}_{CP}(b\ra s\ell^+\ell^-)$ can
nominally constrain such mass-insertion terms. These 
observables should be well-measured at a future flavor factory.
The expected constraints as a function of gluino mass
are shown in Fig.~\ref{fig:supersymmetry1}. These plots 
correspond to 50~ab$^{-1}$ of data, and the measurement of
$\Delta M^{}_s$ from hadron collider experiments CDF and D0
has been included. 
The shaded areas show the regions of parameter space that
would be measured non-zero with at least $3\sigma$ statistical
significance. These shaded regions cover about half the 
parameter space. For the restricted case of a light 
gluino ($M^{}_{\tilde{g}}\simle 100$\gevm), all values 
of mass insertions are covered.

\begin{figure}
\includegraphics[height=.40\textheight]{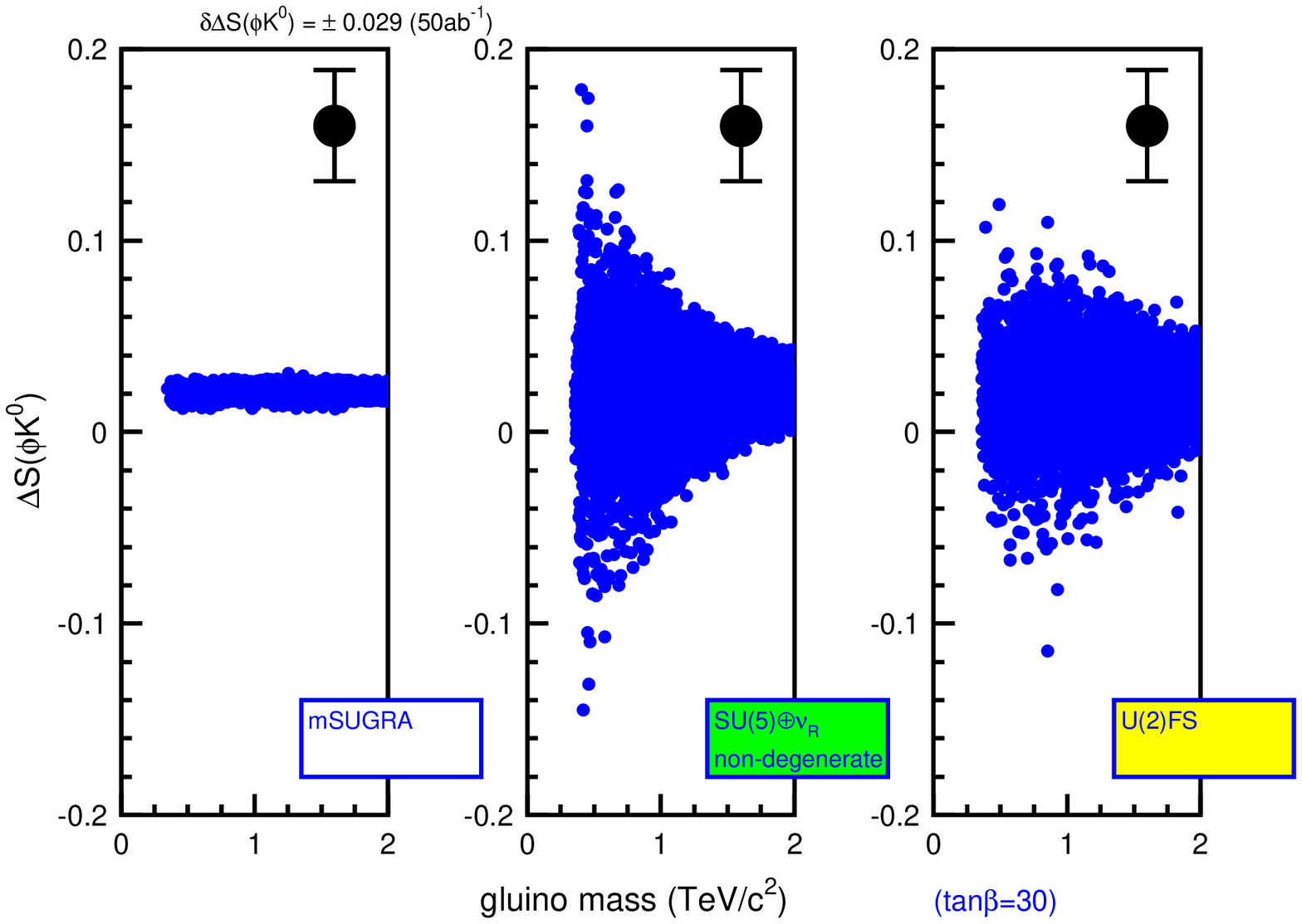}
\caption{Predictions for $\Delta S$ (see text) 
for three supersymmetric theories:
mSUGRA (left), SU(5) SUSY GUT (middle), and MSSM+U(2) (right). 
The data point shown illustrates the error expected from 
50~ab$^{-1}$ of data~\cite{cern_wg_epjc}.}
\label{fig:supersymmetry2}
\end{figure}

\begin{figure}
\includegraphics[height=.40\textheight]{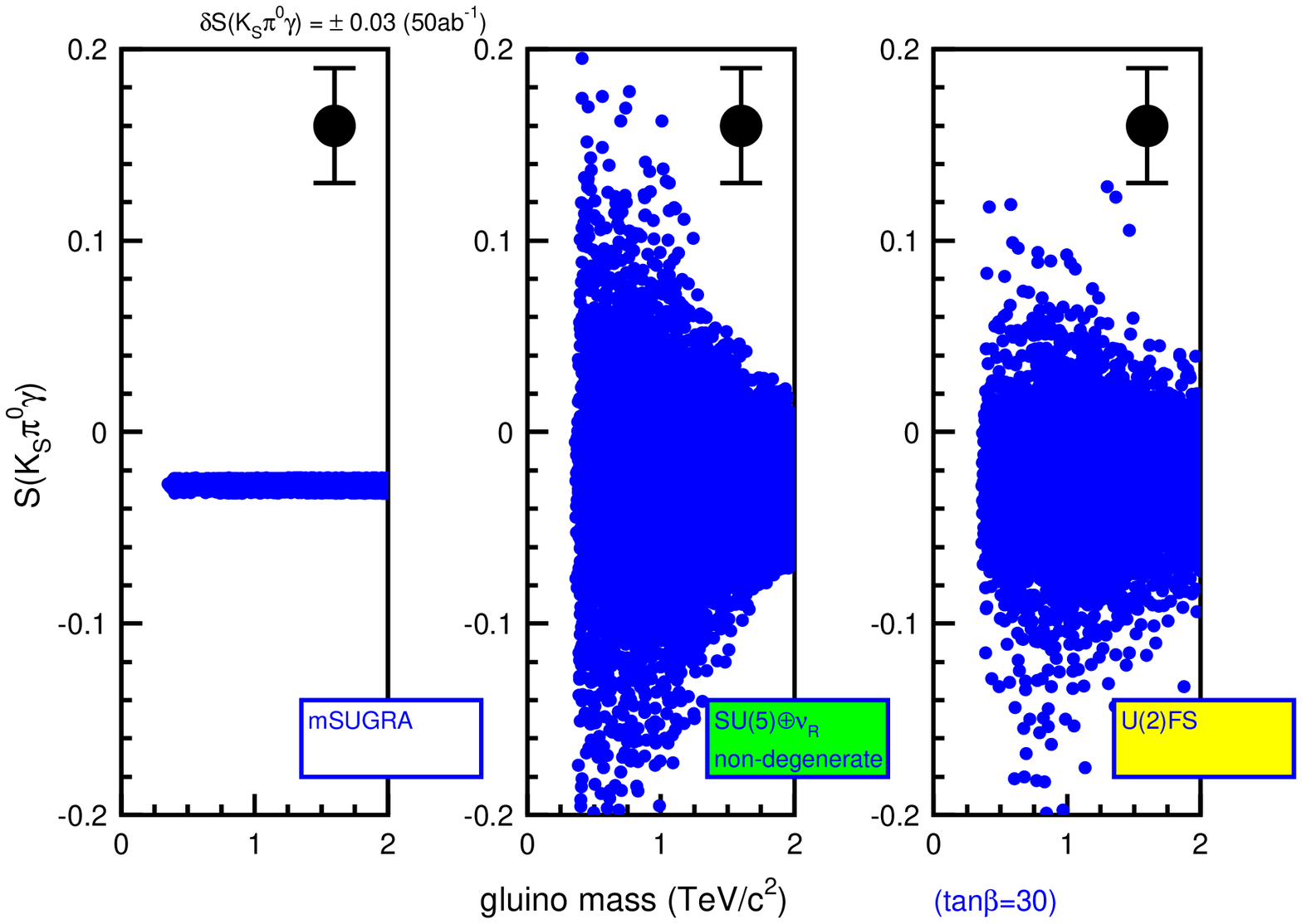}
\caption{Predictions for $(\sin 2\phi^{}_1)_{K^0\pi^0\gamma}$ 
for three supersymmetric theories:
mSUGRA (left), SU(5) SUSY GUT (middle), and MSSM+U(2) (right). 
The data point shown illustrates the error expected from 
50~ab$^{-1}$ of data~\cite{cern_wg_epjc}.}
\label{fig:supersymmetry3}
\end{figure}

A flavor factory can discriminate among different
SUSY-breaking mechanisms via the observables 
$\Delta S = (\sin 2\phi^{}_1)_{q\bar{q}s} - (\sin 2\phi^{}_1)_{c\bar{c}s}$
and $(\sin 2\phi^{}_1)_{K^0\pi^0\gamma}$. 
The results of a Monte Carlo calculation of $\Delta S$ and 
$(\sin 2\phi^{}_1)_{K^0\pi^0\gamma}$ are shown in 
Figs.~\ref{fig:supersymmetry2} and \ref{fig:supersymmetry3},
respectively, for three models of SUSY-breaking: 
mSUGRA, SU(5) SUSY GUT, and MSSM+U(2). For each case 
a spread of points is shown; these result from sampling
over distributions for theoretical parameters whose values 
are unknown. Superimposed on the plots is the error bar
expected from measurements at a flavor factory for
50~ab$^{-1}$ of data. One sees that, depending on the central 
values obtained, a flavor factory could distinguish among the 
models. For example, a central value of 
$(\sin 2\phi^{}_1)_{K^0\pi^0\gamma}>0.03$ would 
essentially rule out mSUGRA, and a large value 
$(\sin 2\phi^{}_1)_{K^0\pi^0\gamma}\approx 0.10$ would favor
SU(5) SUSY GUT with a small gluino mass. 

\section{Summary}

A flavor factory running in the era of the 
LHC would make very large  --\,perhaps decisive\,-- contributions 
to our understanding of beyond-the-SM physics. Such a facility 
would rigorously test our understanding of the SM 
(see Figs.~\ref{fig:weakphase1} and \ref{fig:weakphase2}),
constrain the mass of a charged Higgs 
(Figs.~\ref{fig:bsgamma_theory} and \ref{fig:higgs_vs_tanb}),
measure the values of supersymmetric mass insertion terms
(Fig.~\ref{fig:supersymmetry1}), 
and possibly distinguish among different scenarios of 
supersymmetry breaking 
(Figs.~\ref{fig:supersymmetry2} and \ref{fig:supersymmetry3}).
If supersymmetry is discovered at the LHC, a flavor 
factory may be necessary to determine how it is broken.
In addition, a flavor factory can study $D^0$-$\dbar$ mixing
and search for \cp\ violation in this system; 
a signal for the latter at the percent level would 
be a strong indication of NP. The clean environment of
an $e^+e^-$ flavor factory allows one to search for NP in 
forbidden $\tau^+$ decays such as $\tau^+\ra\mu^+\gamma$.
A future flavor factory is needed to solve the flavor puzzles 
uncovered by the $B$-factory experiments Belle and Babar, 
e.g., $\sin 2\phi^{}_1$ measured in $b\ra sq\bar{q}$
loop processes is systematically lower than that measured
in $b\ra c\bar{c}s$ tree processes; and $A^{}_{CP}$
measured in charged $B^+\ra K\pi$ decays differs substantially
from that measured in neutral $B^0\ra K\pi$ decays.
These measurements are complementary to those that
will be made at LHC experiments, which are based on 
higher-$p^{}_T$ triggers operating in high-multiplicity 
hadroproduction environments. Measurements made at a 
flavor factory may greatly increase our understanding
of results from the LHC.

\begin{theacknowledgments}

We thank the organizers of CIPANP 2009 for 
a stimulating scientific program and excellent hospitality. 
We thank Alexey Petrov and Alex Kagan for reviewing this 
manuscript and suggesting many improvements.

\end{theacknowledgments}

\bibliographystyle{aipproc}   



\end{document}